\documentstyle[12pt]{article}
\begin{document}
\setlength{\parskip}{2ex}
\setlength{\textwidth}{15cm}
\setlength{\textheight}{22.5cm}
\setlength{\oddsidemargin}{0.5cm}
\setlength{\evensidemargin}{0.5cm}
\setlength{\topmargin}{-1cm}
\makeatletter
\@addtoreset{equation}{section}
\makeatother
\renewcommand{\theequation}{\thesection.\arabic{equation}}
\newcommand {\equ}[1] {(\ref{#1})}
\newcommand{\eq}{\begin{equation}}
\newcommand{\eqn}[1]{\label{#1}\end{equation}}
\newcommand{\eea}{\end{eqnarray}}
\newcommand{\eqa}{\begin{eqnarray}}
\newcommand{\eqan}[1]{\label{#1}\end{eqnarray}}
\newcommand{\ba}{\begin{array}}
\newcommand{\ea}{\end{array}}
\newcommand{\eqac}{\begin{equation}\begin{array}{rcl}}
\newcommand{\eqacn}[1]{\end{array}\label{#1}\end{equation}}
\newcommand{\dsx}[1]{{\delta^{\rm S}_{(#1)}}}
\newcommand{\qq}{&\qquad &}
\newcommand{\tfrac}[2]{{\textstyle{\frac{#1}{#2}}}}
\newcommand{\lp}{\left(}\newcommand{\rp}{\right)}
\newcommand{\lc}{\left[}\newcommand{\rc}{\right]}
\newcommand{\lac}{\left\{}\newcommand{\rac}{\right\}}
\newcommand{\pic}{$\spadesuit\spadesuit$}
\def\be{\begin{equation}}
\def\ee{\end{equation}}
\def\bea{\begin{eqnarray}}
\def\eea{\end{eqnarray}}
\def\bean{\begin{eqnarray*}}
\def\eean{\end{eqnarray*}}
\def\ba{\begin{array}} \def\ea{\end{array}}
\def\6{\partial} \def\a{\alpha} \def\b{\beta}
\def\g{\gamma} \def\d{\delta} \def\ve{\varepsilon} 
\def\e{\epsilon}
\def\z{\zeta} \def\h{\eta} \def\th{\theta}
\def\vt{\vartheta} \def\k{\kappa} \def\l{\lambda}
\def\m{\mu} \def\n{\nu} \def\x{\xi} \def\p{\pi}
\def\r{\rho} \def\s{\sigma} \def\t{\tau}
\def\Ph{\phi} \def\ph{\varphi} \def\ps{\psi}
\def\o{\omega} \def\G{\Gamma} \def\D{\Delta}
\def\Th{\Theta} \def\L{\Lambda} \def\S{\Sigma}
\def\PH{\Phi} \def\Ps{\Psi} \def\O{\Omega}
\def\sm{\small} \def\la{\large} \def\La{\Large}
\def\LA{\LARGE} \def\hu{\huge} \def\Hu{\Huge}
\def\ti{\tilde} \def\wti{\widetilde}
\def\non{\nonumber\\}
\def\={\!\!\!&=&\!\!\!}
\def\+{\!\!\!&&\!\!\!+~}
\def\-{\!\!\!&&\!\!\!-~}
\def\id{\!\!\!&\equiv&\!\!\!}
\def\half{\frac{1}{2}}
\renewcommand{\AA}{{\cal A}}
\newcommand{\BB}{{\cal B}}
\newcommand{\CC}{{\cal C}}
\newcommand{\DD}{{\cal D}}
\newcommand{\EE}{{\cal E}}
\newcommand{\FF}{{\cal F}}
\newcommand{\GG}{{\cal G}}
\newcommand{\HH}{{\cal H}}
\newcommand{\II}{{\cal I}}
\newcommand{\JJ}{{\cal J}}
\newcommand{\KK}{{\cal K}}
\newcommand{\LL}{{\cal L}}
\newcommand{\MM}{{\cal M}}
\newcommand{\NN}{{\cal N}}
\newcommand{\OO}{{\cal O}}
\newcommand{\PP}{{\cal P}}
\newcommand{\QQ}{{\cal Q}}
\newcommand{\RR}{{\cal R}}
\newcommand{\SS}{{\cal S}}
\newcommand{\TT}{{\cal T}}
\newcommand{\UU}{{\cal U}}
\newcommand{\VV}{{\cal V}}
\newcommand{\WW}{{\cal W}}
\newcommand{\XX}{{\cal X}}
\newcommand{\YY}{{\cal Y}}
\newcommand{\ZZ}{{\cal Z}}
\def\dti{d\raisebox{0.8em}
{\hspace{-1ex}$\scriptstyle\sim$}{\hspace{0.1em}}}
\def\dxti{dx\raisebox{0.8em}
{\hspace{-1.7ex}$\scriptstyle\sim$}{\hspace{0.1em}}}
\def\Anull{A\raisebox{0.8em}
{\hspace{-1ex}$\scriptstyle\circ$}{\hspace{0.1em}}}
\def\Ati{A\raisebox{0.8em}
{\hspace{-1ex}$\scriptstyle\sim$}{\hspace{0.1em}}}
\def\Anullti{\Ati\raisebox{1.2em}
{\hspace{-1.1ex}$\scriptstyle\circ$}{\hspace{0.1em}}}
\def\Fnull{F\raisebox{0.8em}
{\hspace{-1ex}$\scriptstyle\circ$}{\hspace{0.1em}}}
\def\Fti{F\raisebox{0.8em}
{\hspace{-1ex}$\scriptstyle\sim$}{\hspace{0.1em}}}
\def\Fnullti{\Fti\raisebox{1.2em}
{\hspace{-1.1ex}$\scriptstyle\circ$}{\hspace{0.1em}}}
\def\Dnull{D\raisebox{0.8em}
{\hspace{-1ex}$\scriptstyle\circ$}{\hspace{0.1em}}}
\def\Dti{D\raisebox{0.8em}
{\hspace{-1ex}$\scriptstyle\sim$}{\hspace{0.1em}}}
\def\Dnullti{\Dti\raisebox{1.2em}
{\hspace{-1.1ex}$\scriptstyle\circ$}{\hspace{0.1em}}}
\def\LLnull{$\LL$\raisebox{0.8em}
{\hspace{-1ex}$\scriptstyle\circ$}{\hspace{0.1em}}}
\newcommand{\journal}[4]{{\em #1~}#2\,(19#3)\,#4;}
\newcommand{\aihp}{\journal {Ann. Inst. Henri Poincar\'e}}
\newcommand{\hpa}{\journal {Helv. Phys. Acta}}
\newcommand{\sjpn}{\journal {Sov. J. Part. Nucl.}}
\newcommand{\ijmp}{\journal {Int. J. Mod. Phys.}}
\newcommand{\physu}{\journal {Physica (Utrecht)}}
\newcommand{\pr}{\journal {Phys. Rev.}}
\newcommand{\jetpl}{\journal {JETP Lett.}}
\newcommand{\prl}{\journal {Phys. Rev. Lett.}}
\newcommand{\jmp}{\journal {J. Math. Phys.}}
\newcommand{\rmp}{\journal {Rev. Mod. Phys.}}
\newcommand{\cmp}{\journal {Comm. Math. Phys.}}
\newcommand{\cqg}{\journal {Class. Quant. Grav.}}
\newcommand{\zp}{\journal {Z. Phys.}}
\newcommand{\np}{\journal {Nucl. Phys.}}
\newcommand{\pl}{\journal {Phys. Lett.}}
\newcommand{\mpl}{\journal {Mod. Phys. Lett.}}
\newcommand{\prep}{\journal {Phys. Reports}}
\newcommand{\ptp}{\journal {Progr. Theor. Phys.}}
\newcommand{\nc}{\journal {Nuovo Cim.}}
\newcommand{\app}{\journal {Acta Phys. Pol.}}
\newcommand{\apj}{\journal {Astrophys. Jour.}}
\newcommand{\apjl}{\journal {Astrophys. Jour. Lett.}}
\newcommand{\annp}{\journal {Ann. Phys. (N.Y.)}}
\newcommand{\Nature}{{\em Nature}}
\newcommand{\PRD}{{\em Phys. Rev. D}}
\newcommand{\MNRAS}{{\em M. N. R. A. S.}}
\renewcommand{\title}[1]{\null\vspace{25mm}

\noindent{\Large{\bf #1}}\vspace{10mm}}

\newcommand{\authors}[1]{\noindent{\large #1}\vspace{10mm}

}
\newcommand{\address}[1]{\noindent #1\vspace{10mm}

}
\renewcommand{\abstract}[1]{\vspace{10mm}

\noindent{\small{\em Abstract.} #1}\vspace{2mm}

}

\setcounter{page}{0}
\thispagestyle{empty}
\hspace*{\fill} REF. TUW 96-25

\title{Algebraic characterization of gauge anomalies on a 
nontrivial bundle}

\authors{P. John, 
         O. Moritsch\footnote{Work supported in part by the
         ``\"Osterreichische Nationalbank''
         under Contract Grant Number 5393.}, 
         M. Schweda       
         and S.P. Sorella$^\#$}

\address{Institut f\"ur Theoretische Physik,
         Technische Universit\"at Wien,\\
         Wiedner Hauptstra\ss e 8-10, A-1040 Wien, Austria}
\address{$^\#$ UERJ, Universidade do Estado do Rio de Janeiro, \\
Departamento de F{\'\i}sica Te{\'o}rica, \\
Instituto de F{\'\i}sica,\\
Rua S{\~a}o Francisco Xavier, 524 \\
20550-013, Maracan{\~a}, Rio de Janeiro, Brazil} 
         
\begin{flushleft}
November 1996
\end{flushleft}
\abstract{We discuss the algebraic way of solving the descent equations
corresponding to the BRST consistency condition for the gauge 
anomalies and the Chern--Simons terms on a nontrivial bundle. 
The method of decomposing the exterior derivative as a BRST commutator 
is extended to the present case.}


\newpage


\section{Introduction}

In order to discuss the global structure of the gauge anomalies let us 
recall some properties of the gauge connections defined on a 
principal bundle $P(M,G)$, $M$ being  
an arbitrary  space--time base manifold $M_{2k-2}$ of even 
dimension $(2k-2)$ and  $G$ a compact Lie group\footnote{A general 
introduction to the theory of bundles can be found in 
refs.~\cite{egh,gs}.}.

Connections on $P(M,G)$ are locally represented by Lie algebra 
valued one--forms 
\eq
A = A_\mu dx^\mu = A^a T^a = A^a_\mu dx^\mu T^a \ ,
\eqn{CONNECTION}
where $T^a$ are the antihermitian generators of $G$
\eq
\big[ T^a , T^b \big] = f^{abc} T^c \ ,~~~ Tr(T^a T^b) = \delta^{ab} \ ,
\eqn{REPRESENT}
$f^{abc}$ being the totally antisymmetric structure constants.
The associated two--form field strength is given by
\eq
F=\half F_{\mu\nu} dx^\mu dx^\nu=dA+A^2=dA+\half \big\{ A,A \big\} \ . 
\eqn{FIELD_STRENGTH}
It obeys  the Bianchi identity
\eq
DF=dF+\big[ A,F \big] = 0 \ ,
\eqn{BIANCHI}
where  $D=dx^\mu D_\mu$ is  the covariant exterior derivative with 
respect to the gauge field $A$ and $d=dx^\mu \6_\mu$ is the
nilpotent ordinary exterior space--time derivative.

Gauge transformations of $P(M,G)$ are on $M$ locally represented  
by means of a $G-$valued gauge connection $A_\mu$ transforming as 
\eq
\delta_G A_\mu = -D_\mu \Omega 
= -\6_\mu \Omega - \big[ A_\mu , \Omega \big] \ ,
\eqn{GAUGE_TRANS}
where $\Omega$ is a $G-$valued  infinitesimal gauge parameter.

Infinitesimal diffeomorphisms on $M$ are represented by vector 
fields $v^\mu$. They form an infinite dimensional Lie algebra on $M$,
denoted by Diff($M$). The gauge field transforms under the infinitesimal
diffeomorphisms as a vector, {\it i.e.} 
\eq
\delta_D A_\mu = -L_v A_\mu = -v^\nu \6_\nu A_\mu -(\6_\mu v^\nu)A_\nu \ ,
\eqn{DIFF_TRANS}
where $L_v$ denotes the Lie derivative along the infinitesimal
vector field $v^\mu$. 
The inner derivative $i_v$ with respect to the vector field $v^\mu$
is defined as usual by
\eq
i_v dx^\mu = v^\mu \ .
\eqn{INNER_DERIVATIVE}
As it is well--known, the inner derivative $i_v$, 
the Lie derivative $L_v$ and the exterior derivative $d$ form an 
infinite dimensional graded Lie subalgebra
\eqa
&&\big\{i_v,i_{v'} \big\} = 0 \ ,~~~
\big[L_v,i_{v'} \big] = i_{[v,v']} \ , \non
&&\big\{i_v,d \big\}=L_v \ ,~~~\big[L_v,L_{v'} \big]=L_{[v,v']} \ , \non
&&\big[L_v,d \big] = 0 \ ,~~~~\big\{d,d \big\} = 0 \ ,
\eqan{LIE_SUBALGEBRA}
with $(v,v')~\in$ Diff($M$).

Let us come now to the global properties of a principle bundle. 
A bundle is called trivial if it has the form $P=M \times G$. 
Bundles over contractible base manifolds $M$ are trivial~\cite{egh,gs}.
In the case of a trivial bundle the set of fields introduced so far 
is sufficient in order to describe
the algebraic structure of the gauge anomalies. 
However, as shown in~\cite{sto1}, for the case of a
nontrivial bundle one has to introduce a further gauge field, called 
the reference gauge connection $\Anull$. 
Such a new connection is choosen to be a fixed 
background field, due to the fact that  the action of $v$ on $A$ is  
defined only up to a gauge transformation.
With the help of the reference connection $\Anull$ all vector fields $v$ 
on $M$ can be lifted to $G$.
Again, the infinitesimal gauge transformations form an infinite
dimensional Lie algebra, now denoted by $\tilde g$. Let $U$ be an open
trivializing subset of $M$, $U \subset M$. 
Then the elements of $\tilde g$ are represented on $U$ by pairs of
parameters $(\Omega,v)$
\eq
\Omega \in \Lambda^0(U,g) \ ,~~~v \in \hbox{Diff}(U) \ ,
\eqn{PARAMETERS}
where $\Lambda^0$ denotes the space of zero--forms with values in the 
Lie algebra $g$ of $G$. 
The reference connection is locally represented by the one--form
$\Anull$ on $U$ with values in {g}
\eq
\Anull=\Anull_\mu dx^\mu \in \Lambda^1(U,g) \ .
\eqn{ANULL}
The pairs of parameters obey the commutation relations
\eqa
&&\big[ (\Omega',0),(\Omega,0) \big] 
= ([ \Omega', \Omega ],0) \ , \non
&&\big[ (0,v'),(0,v) \big] = (i_{v'} i_v \Fnull,[v',v]) \ , \non
&&\big[ (0,v),(\Omega,0) \big] = (-L_v\Omega-[i_v \Anull,\Omega],0) \ , 
\eqan{COMMUT}
where $\Fnull$ is the two--form field strength associated to $\Anull$
\eq
\Fnull = \half \Fnull_{\mu\nu} dx^\mu dx^\nu
= d\Anull+\Anull^2 = d\Anull+\half\big\{\Anull,\Anull\big\} \ ,
\eqn{FNULL}
which fulfills the Bianchi identity
\eq
\Dnull\Fnull = d\Fnull + \big[ \Anull , \Fnull \big] = 0 \ .
\eqn{BIANCHI_0}
The reference connection $\Anull$ ensures that the commutators can
be patched together in a consistent way on the overlap $U \cap U'$ of 
two open subsets $U$ and $U'$ of $M$.
This is achieved by replacing the exterior derivative $d$ in the Lie
derivative $L_v$ by the covariant exterior derivative $\Dnull$.
In fact, one has
\eq
\big[ (0,v),(\Omega,0) \big] = (-\hbox{\LLnull}_v\Omega,0) \ ,
\eqn{LIE_0}
with $\hbox{\LLnull}_v = i_v \Dnull + \Dnull i_v$. Note that the
ordinary Lie derivative $L_v$ in \equ{DIFF_TRANS} can be rewritten as
$L_v=i_v d + d i_v$.

Different reference connections $\Anull$ on $P$ yield isomorphic
Lie algebras $\tilde g$. Note that the pure gauge transformations
$(\Omega,0)$ form an ideal of $\tilde g$, while the transformations
with respect to the vector fields $(0,v)$ in general do not form
a subalgebra of $\tilde g$.

Moreover, if the bundle $P$ is trivial, one can choose $U=M$ and
$\Anull=0$. Then $\tilde g$ is the semi--direct product of
$\Lambda^0(M,g)$ and Diff($M$), meaning that every 
nontrivial bundle is locally trivial. 

The affine space of all connections on $P$ carries an affine 
representation $\RR$ of $\tilde g$ given locally by
\eq
\RR(\Omega,v)A=-D\big(\Omega+i_v(A-\Anull)\big) - i_vF \ ,
\eqn{RR}
where $A,\Anull \in \Lambda^1(U,g)$~\footnote{$\Lambda^1(U,g)$ denotes  
the space of the one--forms with values in the Lie algebra $g$.} 
are the local expressions on $U$
of the connections on $P$ and $D$ is the covariant exterior derivative 
with respect to $A$.
Finally, by definition, the fixed reference connection $\Anull$ does 
not transform under $\tilde g$
\eq
\RR(\Omega,v) \Anull = 0 \ .
\eqn{RANULL}


\section{The BRST transformations}

In order to introduce the nilpotent BRST operator $s$  let us replace, 
as usual, the infinitesimal parameters $(\Omega,v)$ introduced in the 
previous section with Grassmann ghost fields. 
We will introduce therefore two ghost fields, $c$ and $\xi^\mu$, 
associated to the gauge and to the diffeomorphism 
transformations, respectively. 
Both ghosts have form degree zero and ghost number one.

Concerning the BRST transformations~\cite{brst} of the various 
fields and ghosts, we proceed by making use of the so--called
Maurer--Cartan horizontality conditions~\cite{tmieg,bau}. 
For this purpose we define the nilpotent
differential operator of total degree one\footnote{The total degree 
is defined as the sum of the form degree and of the ghost number.}
\eq
\dti = d-s \ ,~~~\dti^2=0 \ ,
\eqn{d_TILDE}
where $d$ is the exterior derivative.
Furthermore, we define the generalized gauge 
connection~\cite{tmieg,bau} 
according to
\eq
\Ati = A + c + i_\xi(A-\Anull) \ ,
\eqn{A_TILDE}
with $i_\xi$ as the inner derivative with respect to the diffeomorphism 
ghost $\xi^\mu$
\eq
i_\xi dx^\mu=\xi^\mu \ .
\eqn{inner-xi}
The generalized reference connection is simply given by
\eq
\Anullti = \Anull \ ,
\eqn{ANULL_TILDE}
expressing the fact that $\Anull$ is a fixed background field which 
does not transform.

The corresponding generalized field strengths of total degree two are 
given by
\eq
\Fti=\dti \Ati +\Ati^2 ~~~\hbox{and}~~~
\Fnullti=\dti \Anullti + \Anullti^2 \ ,
\eqn{STRENGTH_TILDE}
and obey the generalized Bianchi identities
\eq
\Dti\Fti=\dti\Fti+\big[ \Ati,\Fti \big]=0 ~~~\hbox{and}~~~
\Dnullti\Fnullti=\dti\Fnullti+\big[ \Anullti,\Fnullti \big]=0 \ ,
\eqn{BIANCHI_TILDE}
with $\Dti$ and $\Dnullti$  the generalized covariant exterior 
derivatives with respect to $\Ati$ and to $\Anullti$. 

The Maurer--Cartan horizontality conditions can then be expressed by
\eqa
\Fti \= \half F_{\mu\nu} \dxti^\mu \dxti^\nu \ , \non
\Fnullti \= \half \Fnull_{\mu\nu} \dxti^\mu \dxti^\nu \ , 
\eqan{MCHC}
with $\dxti^\mu$ the generalized  differential of degree one 
\eq
\dxti^\mu=dx^\mu+\xi^\mu \ . 
\eqn{DX_TILDE}

Expanding these conditions according to the ghost number and the form 
degree one gets the following set of nilpotent BRST transformations  
\eqa
sc\=c^2-\hbox{\LLnull}_\xi c + \half i_\xi i_\xi \Fnull \ , \non
sA\=D\big(c+i_\xi(A-\Anull)\big)-i_\xi F \ , \non
s\Anull\=0 \ , \non
sF\=\big[\big(c+i_\xi(A-\Anull)\big),F\big]-\LL_\xi F \ , \non
s\Fnull\=0 \ ,
\eqan{BRST_SET}
where $\LL_\xi=i_\xi D -D i_\xi$ and
$\hbox{\LLnull}_\xi=i_\xi \Dnull -\Dnull i_\xi$ are 
the covariant Lie derivatives with respect to $A$ and $\Anull$,
respectively\footnote{Notice that the relative sign
in the definition of the Lie derivatives has now been changed due to the
fact that $\xi$ is an odd variable, carrying ghost number one.
This implies that $i_\xi$ is even and that $\LL_\xi$, 
$\hbox{\LLnull}_\xi$ have odd degree.}.
The above BRST transformations are completed by giving the 
transformation of the diffeomorphism ghost
\eq
s\xi^\mu=-\xi^\nu\6_\nu\xi^\mu \ .
\eqn{BRS_DIFF}

Instead of the ghost $c$ it turns out to be more useful to use the 
following combination
\eq
\hat c = c + i_\xi(A-\Anull)  \ ,
\eqn{C_HAT}
as the basic variable. In terms of the new variable  $\hat c$, 
the BRST transformations read now 
\eqa
s\xi^\mu\=-\xi^\nu\6_\nu\xi^\mu \ , \non
s\hat c\=\hat{c}^2 - \hat F \ , \non
sA\=D\hat c - i_\xi F \ , \non
s\Anull\=0 \ , \non
sF\=\big[\hat c ,F\big]-\LL_\xi F \ , \non
s\Fnull\=0 \ ,
\eqan{BRST_SET2}
with $\hat F$ defined as  
\eq
\hat F = \half i_\xi i_\xi F \ .
\eqn{F_HAT}
Notice, finally, that $\hat F$ has form degree zero, ghost number 
two and that it transforms covariantly, {\it i.e.} 
\eq
s\hat F = \big[ \hat c , \hat F \big] \ .
\eqn{BRS_FHAT}


\section{Decomposition formula and algebraic relations}

Before going further, let us make some comments on the 
functional space which we shall use in the following. 
As already done in previous works~\cite{sor1,mo1}, we shall assume that 
the functional space the BRST operator will act upon is identified 
with the space of form--polynomials constructed out of the variables 
$(\xi^\mu, c, A -\Anull)$ and the differentials 
$(d\xi^\mu, dc, dA, d\Anull)$. Of course, on the local space of 
form--polynomials this set of variables can always
be replaced by the equivalent basis given by 
$(\xi^\mu,d\xi^\mu,\hat c,\Dnull \hat c,(A-\Anull),F,\Fnull)$. 
Due to the presence of the covariant derivative $\Dnull$ and of 
the curvatures $(F,\Fnull)$ 
one can easily understand the latter choice, which
turns out to be more convenient to solve the BRST consistency condition. 
We remark also that the use of the combination $(A-\Anull)$  
steams from the fact that it  emerges rather naturally in the 
BRST transformations \equ{BRST_SET2} and in the formula \equ{C_HAT}.  
Let us recall, finally, that the space of form--polynomials is the 
most suitable choice in order to discuss the anomalies and the 
Chern--Simons terms which, as it is well--known, can be written in 
terms of differential forms.
     
Let us now introduce an operator $\d$ of total degree zero 
defined in the following way~\cite{sor1,mo1}
\eqa
\d\xi^\mu =-dx^\mu \ ,~~~
\d\ph=0~~~{\rm for}~~~\ph=(c , A , \Anull , F , \Fnull) \ .
\eqan{DECOMP}
>From the above expression one sees that $\d$ increases the form 
degree and decreases the ghost number by one unit respectively and 
that it only on the diffeomorphism ghost acts.
The usefulness of the operator $\d$ relies on the fact that it 
allows to decompose
the covariant exterior space--time derivative $\Dnull$ as a 
BRST commutator. In fact, we have 
\eq
[s,\d]=-\Dnull \ .
\eqn{BRST_COMMUT}
Let us remark that the presence of the covariant derivative 
$\Dnull$ in eq.\equ{BRST_COMMUT} makes all the difference with 
respect to the case of a trivial bundle, where  only the ordinary 
exterior space--time derivative $d$ appears. 
The operators $(s, \d, \Dnull)$ give rise to the following 
algebraic relations: 
\eqa
\big[ s,\d\big]\=-\Dnull \ ,~~~\big[\Dnull,\d\big]
=\big[ d,\d\big]=2\GG \ , \non[3mm]
\big\{ s,\Dnull\big\}\=\big\{ s,d\big\}=s^2=d^2=0 \ ,~~~
\big[\GG,\d\big]=0 \ , \non[3mm]
\big\{\GG,\Dnull\big\}\=\big\{\GG,d\big\}=0 \ ,~~~\GG\GG=0 \ , \non[3mm]
\big\{\GG,s\big\}\=\KK=\Dnull\Dnull \ ,~~~
\big[ \KK, s \big] = 0 \ , \non[3mm]
\big[ \KK,\Dnull \big] \=\big[ \KK,d \big]=0 \ , \non[3mm]
\big[ \KK, \d \big] \=0 \ ,~~~\big[ \KK,\GG \big] =0 \ ,
\eqan{ALGEBRA}
where the operator $\GG$ is defined as 
\eq
\GG \hat c=\Fnull \ ,~~~
\GG\ph=0~~~{\rm for}~~~
\ph=(\xi^\mu,d\xi^\mu,\Dnull \hat c,(A-\Anull),F,\Fnull) \ .
\eqn{GG} 
It decreases the ghost number by one unit and increases the 
form degree by two units, hence it is an operator of total degree one.

Let us also note that in the case of a trivial bundle $(\Anull=0)$ the
operator $\GG$ is absent. This is consistent with the results found
in~\cite{mo1}. In other words, its presence is related to the global 
properties of the bundle.


\section{Cohomology and descent equations}

It is well--known that the search for the invariant Lagrangians and
the anomalies corresponding to a given set of field
transformations can be done in a purely algebraic way by solving
the BRST consistency conditions in the space of the
integrated local field polynomials.

\noindent
This leads to the study of the nontrivial solutions of the following
cohomology problem
\eq
s\D=0~~~,~~~\D \not= s\bar{\D} \ ,
\eqn{CE}
where $\D$ and $\bar{\D}$ are integrated
local field polynomials.
Setting $\D=\int\QQ^g_n$, the condition \equ{CE} translates at 
the nonintegrated level as  
\eq
s\QQ^g_n+d\QQ^{g+1}_{n-1}=0 \ ,
\eqn{LE}
where $\QQ^g_n$ is some local polynomial in the fields
with ghost number $g$ and form degree $n$, $n$ denoting the 
dimension of the space--time. $\QQ^g_n$ is said to be nontrivial if
\eq
\QQ^g_n \not= s\bar\QQ^{g-1}_{n}+d\bar\QQ^{g}_{n-1} \ .
\eqn{NONTRIVIAL}
In this case the integral of $\QQ^g_n$ on space--time 
identifies a cohomology class of the BRST operator $s$ and, 
according to its ghost number, it corresponds to an invariant 
Lagrangian $(g=0)$ or to an anomaly $(g=1)$.

The nonintegrated  equation \equ{LE}, due to the algebraic 
Poincar\'e Lemma~\cite{cotta,dragon}, is easily seen to generate 
a tower of descent equations
\eqa
&&s\QQ^{g}_{n}+d\QQ^{g+1}_{n-1}=0 \ , \non
&&s\QQ^{g+1}_{n-1}+d\QQ^{g+2}_{n-2}=0 \ , \non
&&~~~~~.....\non
&&~~~~~.....\non
&&s\QQ^{g+n-1}_{1}+d\QQ^{g+n}_{0}=0 \ , \non
&&s\QQ^{g+n}_{0}=0 \ .
\eqan{LADDER}

As it has been well--known for several years, that these equations can be
solved by using a transgression procedure based on the so--called
{\it Russian formula}
\cite{sto1,tmieg,bau,witt,dvio,band,gins,tonin1,sto2,brandt}.
More recently an alternative way of finding nontrivial solutions of
the ladder \equ{LADDER} has been proposed and
successfully applied in the study of the Yang--Mills gauge
anomalies~\cite{sor1}.
The method makes use of the decomposition formula \equ{BRST_COMMUT}.
One easily verifies that, once the decomposition has
been found, successive applications of the operator $\d$ on the
zero--form $\QQ^{g+n}_{0}$ which solves the last equation of the tower
\equ{LADDER} give an explicit nontrivial solution for the higher
cocycles.

It is a remarkable fact that solving the last equation of the
tower of descent equations \equ{LADDER} is only a problem of local 
BRST cohomology instead of a modulo--$d$ one.
One sees then that, due to the operator $\d$, the study of the
cohomology of $s$ modulo $d$ is essentially
reduced to the study of the local cohomology of $s$. The latter   
can be {\em e.g.} systematically analyzed by using the powerful 
technique of the spectral sequences~\cite{dixon}.


\section{Chern--Simons terms and gauge anomalies}

In order to solve the descent equations \equ{LADDER}
we use the following strategy.  First, we look at the
general nontrivial solution of the last descent equation. Then,
by using the operators $\d$ and $\GG$ with the help of the 
algebra \equ{ALGEBRA}, we solve the tower iteratively as done 
in~\cite{sor1}. 

Let $n=2k$ be the dimension of the base manifold $M$
of a nontrivial bundle $P(M,G)$ with structure group $G$.
Introducing a covariant BRST operator according to
\eq
S = s - \hat c \ ,
\eqn{BRST_COV}
one obtains a remarkable correspondence between the 
form sector and the gauge sector of the theory
\eqa
DF=dF+\big[A,F\big]=0 \ ,~~~S\hat F=s\hat F-\big[\hat c,\hat F\big] \ .
\eqan{CORRESP}
Let us define now the interpolating shifted gauge ghost
\be
\hat c(t)=t\hat c~~~,~~~t\in[0,1] \ ,
\ee
with $\hat c(0)=0$ and $\hat c(1)=\hat c$,
and the associated ghost field strength
\be
\hat F(t)=-s\hat c(t)+\hat c(t)\hat c(t) \ ,
\ee
with $\hat F(0)=0$ and $\hat F(1)=\hat F$.
With the help of the interpolating generalized covariant BRST operator  
\be
S_t=s-\hat c(t) \ ,
\ee
with $S_0=s$ and $S_1=S$,
one finds the following identities
\bea
\frac{d\hat F(t)}{dt}=-S_t\hat c~~~,~~~S_t\hat F(t)=0 \ .
\eea
Therefore, in a space--time with dimension $2k$ one has
\bea
Tr(\hat F^k)\=Tr\big(\hat F^k(1)-\hat F^k(0)\big)
=Tr\int^1_0dt\,\frac{d}{dt}\,\hat F^k(t) \non
\=k\,Tr\int^1_0dt\,\frac{d\hat F(t)}{dt}\,\hat F^{k-1}(t)
=-k\,Tr\int^1_0dt\,(S_t\hat c)\hat F^{k-1}(t) \non
\=-s\big(k\,Tr\int^1_0dt\,\hat c\,\hat F^{k-1}(t)\big) \ .
\eea
Using the nilpotency of the BRST operator and the 
fact that $Tr(\hat F^k)\neq 0$ in a space--time with dimension $2k$, 
one may conclude that $k\,Tr\int^1_0dt\,\hat c\,\hat F^{k-1}(t)$ is 
nontrivial.
Since $Tr(\hat F^k)$ contains the product of $2k$ fermionic 
diffeomorphism ghosts $\x^\m$, it follows:

\begin{quote}
{\em In a space--time with dimension $2k-1$ a 
nontrivial solution of the local equation $s\QQ^{2k-1}_0=0$
on a nontrivial bundle
can be represented by the integrated parametric formula}
\be
\QQ^{2k-1}_0=-k\,Tr\int^1_0dt\,\hat c\,\hat F^{k-1}(t) \ .
\label{IPF}
\ee
\end{quote}

Acting with the operator $\GG$ on the last of the descent 
equations \equ{LADDER} 
and using the algebra \equ{ALGEBRA} one finds
\eq
\GG s \QQ^{2k-1}_0 = 0 = -s\GG\QQ^{2k-1}_0 + \Dnull\Dnull\QQ^{2k-1}_0 \ ,
\eqn{GG_BRST_CLOSED}
which, due to the vanishing of the last term in \equ{GG_BRST_CLOSED}, 
implies that $\GG\QQ^{2k-1}_0$ is BRST--closed
\eq
s\GG\QQ^{2k-1}_0 = 0 \ .
\eqn{GG_CLOSED}
The general solution of \equ{GG_CLOSED} has the form
\eq
\GG\QQ^{2k-1}_0 = s\bar\QQ^{2k-3}_2 +\hat\QQ^{2k-2}_2 \ ,
\eqn{GG_SOL}
with a possible nontrivial part $\hat\QQ^{2k-2}_2$
\eq
s\hat\QQ^{2k-2}_2 = 0 \ ,~~~\hat\QQ^{2k-2}_2 \not= s\hat\QQ^{2k-3}_2 \ .
\eqn{NONTRIV}
Recalling  that the operator $\GG$ acting on the variable 
$\hat c$ yields the field $\Fnull$, we infer that 
\eq
\hat\QQ^{2k-2}_2 = Tr\big(\Fnull \PP (\hat c,\hat F)\big)
= \Fnull^a \PP^{a} (\hat c,\hat F) \ ,
\eqn{QQ_HAT}
with $\PP$ some polynomial in the fields $\hat c$ and $\hat F$ 
having form degree zero and ghost number $2k-2$. 

In the following we will show that $\GG\QQ^{2k-1}_0$ is not only
BRST--closed but also BRST--exact
\eq
\GG\QQ^{2k-1}_0 = s\bar\QQ^{2k-3}_2 \ .
\eqn{GG_EXACT}
The exactness of $\GG\QQ^{2k-1}_0$ will be proven by 
showing that the polynomial ${\PP^{a}}$ is BRST trivial, 
due to the fact that the variable $\Fnull$ is BRST invariant. 
Recalling then that ${\PP^a}$ depends only on 
$(\hat c,\hat F)$ due to the fact that it has form 
degree zero, let us first express the rigid gauge 
transformations of $(\hat c,\hat F)$ as BRST anticommutators, 
{\it i.e.}\footnote{Here we have taken into account only the 
part of the BRST operator relative to the fields $\hat c,\hat F$.}
\eq
\big\{s,\frac{\6}{\6\hat c^a}\big\} = \d^a_{rig.} 
= f^{abc} \hat c^b \frac{\6}{\6 \hat c^c}
+ f^{abc} \hat F^b \frac{\6}{\6 \hat F^c} \ ,
\eqn{BRST_ANTI}
It is easily checked by using the Jacobi identity that the 
operator $\d^a_{rig.}$ commutes with the BRST operator 
\eq
\big[\d^a_{rig.}, s\big]=0 \ .
\eqn{RIG_COMM}
Furthermore, one has
\eq
\d^a_{rig.} \Fnull^b = 0 \ .
\eqn{RIG_FNULL}
Decomposing now the eq.\equ{QQ_HAT} in terms of the 
eigenvalues of $\d^a_{rig.}$, we get  
\eq
\d^a_{rig.} \big( \Fnull^b \PP^b \big)_{(l)} 
= l \big( \Fnull^b \PP^b \big)_{(l)} \ ,~~~
\Fnull^b \PP^b=\sum_l \big( \Fnull^b \PP^b \big)_{(l)} \ . 
\eqn{EVP}
For $l\not=0$ one obtains
\eqa
\big( \Fnull^b \PP^b \big)_{(l)} 
\= \frac{1}{l} \d^a_{rig.} \big( \Fnull^b \PP^b \big)_{(l)} \non 
\= \frac{1}{l} \big\{s,\frac{\6}{\6\hat c^a}\big\} 
\big( \Fnull^b \PP^b \big)_{(l)} \non
\=s \Big( \frac{1}{l} \frac{\6}{\6\hat c^a} 
\big( \Fnull^b \PP^b \big)_{(l)}\Big) \ , 
\eqan{LNOT}
since from eq.\equ{NONTRIV} and eq.\equ{EVP} it follows that  
\eq
s\big( \Fnull^b \PP^b \big)_{(l)} = 0 \ .
\eqn{BRS_NULL}
However, contributions of the type \equ{LNOT} 
can be neglected, since they are BRST trivial. 
What remains is the term corresponding  to the eigenvalue $l=0$ 
\eq
\d^a_{rig.} \big( \Fnull^b \PP^b \big)_{(0)} = 0 \ . 
\eqn{LZERO}
The above equation states that $\big(\Fnull^b \PP^b \big)_{(0)}$ 
has to be invariant 
under rigid gauge transformations. However, since $\Fnull$ is 
left invariant 
by $\d^a_{rig.}$, eq.\equ{LZERO} implies that the 
polynomial $\PP(\hat c,\hat F)$ has to be invariant as well. 
It follows therefore that $\big( \Fnull^b \PP^b \big)_{(0)}$ vanish. 
This completes our proof.

Repeating and generalizing these arguments to higher 
$\GG$--cocycles\footnote{In this case one has to incorporate in the rigid
gauge transformations of \equ{BRST_ANTI} also the contributions of 
the remaining fields with nonvanishing form degree.}
one sees that the identity \equ{GG_EXACT} generates a subtower of 
descent equations between the operators $s$ and $\GG$
\eqa
&&\GG\QQ^{2k-1}_0 = s\bar\QQ^{2k-3}_2 \ , \non
&&\GG\bar\QQ^{2k-3}_{2} = s\bar\QQ^{2k-5}_{4} \ , \non
&&~~~~~.....\non
&&~~~~~.....\non
&&\GG\bar\QQ^{3}_{2k-4} = s\bar\QQ^{1}_{2k-2} \ , \non
&&\GG\bar\QQ^{1}_{2k-2} = \PP^{0}_{2k}(\Fnull) \ , 
\eqan{SUBTOWER}
which end up with a  polynomial   $\PP^{0}_{2k}(\Fnull)$
of ghost number zero and form degree $2k$. Moreover, from eq.\equ{GG} it 
follows that $\PP^{0}_{2k}(\Fnull)$ is nothing else than 
the Chern form of $\Fnull$ in $2k$ space--time dimensions.

Equiped with this algebraic setup we are now ready  to discuss the 
solutions of the descent equations \equ{LADDER}.
Before  analyzing the general case of a  space--time of 
dimension $n=2k-1$ let us first solve, as an explicit example,  
the three--dimensional case $(k=2)$
\eqa
&&s\QQ^0_3+d\QQ^1_2=0 \ , \non
&&s\QQ^1_2+d\QQ^2_1=0 \ , \non
&&s\QQ^2_1+d\QQ^3_0=0 \ , \non
&&s\QQ^3_0=0 \ .
\eqan{TOWER_3} 

\noindent
The nontrivial solution of the last descent equation follows 
from \equ{IPF} with $(k=2)$
\eq
\QQ^{3}_0 = -2\,Tr\int^1_0dt\,\hat c\,\hat F(t) 
= Tr\big(\frac{1}{3}\hat c^3 -\hat F\hat c \big) \ .
\eqn{SOL_3DIM} 
Acting with the operator $\d$ on the last equation one gets
\eq
\big[\d,s\big]\QQ^3_0 + s\d\QQ^3_0 = 0 \ ,
\eqn{COC}
which, using the decomposition \equ{BRST_COMMUT}, 
becomes\footnote{Remark that all polynomials
$\QQ$ are gauge invariant and therefore the covariant
exterior derivative $\Dnull$ reduces to the ordinary one.}
\eq
s\big(\d\QQ^3_0\big) + d\QQ^3_0 = 0 \ .
\eqn{COCYCLE1}
This equation shows that $\d\QQ^3_0$ provides  a solution for the  
cocycle $\QQ^2_1$ in eqs.\equ{TOWER_3}.

In order to find an expression for the next cocycle, 
we apply  the operator $\d$ on the eq.\equ{COCYCLE1}. 
Using then the algebraic relations \equ{ALGEBRA}, one has
\eq
s\big(\frac{\d^2}{2}\QQ^3_0\big)-\GG\QQ^3_0+d\big(\d\QQ^3_0\big) = 0 \ .
\eqn{COCYCLE2}
According to \equ{GG_EXACT}, the term $\GG\QQ^3_0$ in \equ{COCYCLE2} 
can be rewritten as
\eq
\GG\QQ^3_0 = s Tr \big(\Fnull \hat c\big) = s \bar\QQ^1_2  \ ,
\eqn{GGQQ3}
so that eq.\equ{COCYCLE2} becomes 
\eq
s\big(\frac{\d^2}{2}\QQ^3_0-\bar\QQ^1_2\big)+d\big(\d\QQ^3_0\big) = 0 \ .
\eqn{COCYCLE2A}

Repeating again the previous steps,  for the highest level we obtain 
\eq
s\big(\frac{\d^3}{3!}\QQ^3_0-\d\bar\QQ^1_2\big)
+d\big(\frac{\d^2}{2}\QQ^3_0-\bar\QQ^1_2\big) = 0 \ ,
\eqn{COCYCLE3}
from which it follows that $\QQ^0_3$ can be identified 
with $(\frac{\d^3}{3!}\QQ^3_0-\d\bar\QQ^1_2)$, showing then 
the usefulness of the  operators $\d$ and $\GG$ in solving the 
tower \equ{TOWER_3}.  

Summarizing, the algebraic solution
of the three--dimensional tower of descent equations \equ{TOWER_3}
in the case of a nontrivial bundle is given by
\eqa
\QQ^0_3\=\frac{\d^3}{3!}\QQ^3_0-\d\bar\QQ^1_2 \ , \\
\QQ^1_2\=\frac{\d^2}{2}\QQ^3_0-\bar\QQ^1_2 \ , \\
\QQ^2_1\=\d\QQ^3_0 \ , 
\eqan{ALG3}
with
\eqa
\QQ^3_0\=Tr\big(\frac{1}{3}\hat c^3 -\hat F \hat c\big) \ , \\
\QQ^2_1\=Tr\big(-(A-\Anull)\hat c^2 +(A-\Anull)\hat F
+i_\xi F \hat c\big) \ , \\
\QQ^1_2\=Tr\big((A-\Anull)^2\hat c -(A-\Anull)i_\xi F 
-(F+\Fnull)\hat c\big) \ , \\
\QQ^0_3\=Tr\big(-\frac{1}{3}(A-\Anull)^3 +(A-\Anull)(F+\Fnull)\big) \ ,
\eqan{COC3}
In particular, $\QQ^0_3$ is recognized to be the  three--dimensional 
Chern--Simons form, while $\QQ^1_2$ yields the  two--dimensional  
gauge anomaly which, by using the definition
of the shifted gauge ghost \equ{C_HAT}, can be rewritten as 
\eq
\QQ^1_2=-Tr\big(c\,\Dnull A +c\,d\Anull+ \Fnull i_\xi (A-\Anull) \big) \ .
\eqn{ANOM2}
In the case of a trivial bundle $(\Anull=0)$ the anomaly \equ{ANOM2}
reduces to the well--known familiar expression 
\eq
\QQ^1_2=-Tr\big(c\,dA) \ .
\eqn{ANOM2_TRIVIAL}
Finally, acting with the exterior derivative $d$ on $\QQ^0_3$ one obtains
the related Chern form
\eq
d\QQ^0_3 = Tr \big(F^2-\Fnull^2\big) \ .
\eqn{CHERN4}

Let us come now to the general case of a space--time with 
dimension $n=2k-1$, 
\eqa
&&s\QQ^{0}_{2k-1}+d\QQ^{1}_{2k-2}=0 \ , \non
&&s\QQ^{1}_{2k-2}+d\QQ^{2}_{2k-3}=0 \ , \non
&&~~~~~.....\non
&&~~~~~.....\non
&&s\QQ^{2k-2}_{1}+d\QQ^{2k-1}_{0}=0 \ , \non
&&s\QQ^{2k-1}_{0}=0 \ ,
\eqan{LADDER_GENERAL}
It is straightforward to iterate the previous construction to obtain
the solution of the descent equations \equ{LADDER_GENERAL}. 
The latter is given by 
\eqa
\QQ^{2k-1}_0\=-k\,Tr\int^1_0dt\,\hat c\,\hat F^{k-1}(t) \ , \non
\QQ^{2k-1-2p}_{2p}\=\frac{\d^{2p}}{(2p)!}\QQ^{2k-1}_0
-\sum_{q=0}^{p-1}\frac{\d^{2q}}{(2q)!}\bar\QQ^{2k-1-2p+2q}_{2p-2q} \ ,
\eqan{COC_EVEN}
for the even space--time sector and
\eqa
\QQ^{2k-2}_1\=\d\QQ^{2k-1}_0 \ , \non
\QQ^{2k-2-2p}_{2p+1}\=\frac{\d^{2p+1}}{(2p+1)!}\QQ^{2k-1}_0
-\sum_{q=0}^{p-1}\frac{\d^{2q+1}}{(2q+1)!}\bar\QQ^{2k-1-2p+2q}_{2p-2q} \ ,
\eqan{COC_ODD}
for the odd space--time sector and $p=1,2,...,k$.
The solutions \equ{COC_EVEN} and \equ{COC_ODD} generalize the results
of~\cite{sor1} to the case of a nontrivial bundle.



\end{document}